\pdfoutput=1
\documentclass[a4paper]{tufte-handout}
\usepackage{graphicx}
\setkeys{Gin}{width=\linewidth,totalheight=\textheight,keepaspectratio}
\graphicspath{{Figures/}}
\usepackage{booktabs}
\usepackage{units}
\usepackage{fancyvrb}
\fvset{fontsize=\normalsize}
\usepackage{multicol}
\usepackage{amsfonts}
\usepackage{natbib}
\usepackage{microtype}
\def\adx#1:#2\par{\par\halign{\hskip #1##\hfill\cr #2}\par}

\def\teff{T_{\rm eff}}
\def\mast{M_\ast}

\def\msol{M_\odot}

\def\last{L_\ast}
\def\mmsol{M/M_\odot}
\def\llsol{L/L_\odot}

\def\nabrad{\nabla_{\mathrm{rad}}}

\def\H1{^1\mathrm{H}}
\def\He3{^3\mathrm{He}}
\def\He4{^4\mathrm{He}}
\def\C12{^{12}\mathrm{C}}
\def\N14{^{14}\mathrm{N}}
\def\O16{^{16}\mathrm{O}}
\def\diff{{\mathrm d}}
\def\Diff{{\mathrm D}}
\def\unity{ \hbox{1\kern-.23em l} }
\def\zero{ \hbox{0\kern-.23em |} }
\def\field{ \hbox{I\kern-.23em K} }

\def\tens#1{\mathbb{#1}}

\def\braket #1.#2.{\langle #1 \vert #2 \rangle}

\def\O{\mathcal{O}}
\usepackage[]{microtype}

\newcommand{\lyxaddress}[1]{
\vspace{1.4em}
\par {\raggedright #1
\vspace{1.4em}
\noindent\par}
}

\usepackage{amsmath} 

\title{The theorem that was none - I. Early history}
\author{Alfred Gautschy}
\date{  }  
\begin{document}
\maketitle  % prints here title, author, and date  

\lyxaddress{CBmA Liestal \& ETH-Bibliothek, R\"amistrasse 101, 8092
  Z\"urich, Switzerland}

\begin{abstract}
\noindent The early history of the Vogt-Russell theorem is retraced following
its route starting at the realization of a correlation between mass and
luminosity of binary and pulsating stars, through the embossing of this
observation into a theorem, and finally to the emerging first signs of
its failure to serve as a theorem in the strict mathematical sense of the word. 
\end{abstract}

\section*{Introduction}

Astrophysics is not an exact science in the sense of mathematics.
On the observational side, time and again the Universe is considerably
more creative than human imagination; on the theoretical side, the equations
that are derived to model the intricate processes and phenomena in
astrophysics usually lack the simple symmetries or properties to
attract mathematicians' attention for in-depth analyses of their properties.
Therefore, proclaiming a theorem in astrophysics~--~by astrophysicists~--~was, 
is, and likely will remain a daring undertaking. 

\newthought{THE theorem in stellar physics}, the
topic of this marginalia, is known by the name of Vogt-Russell (VR) theorem:
It claims, roughly speaking, that the structure of any
star is uniquely determined by its mass and its chemical composition alone. 
Today we know that the theorem does not hold in its original strict formulation; nonetheless, the VR theorem continues to enjoy some popularity and 
pops up in almost all courses on stellar structure and evolution, 
in textbooks, and even recently in research papers. 
%---------------------------------------------------------------------------
\sidenote{e.g. \citet{Carroll2013} or \,\,\,\,\,\, \citet{Melis2014}}
%---------------------------------------------------------------------------
The following exposition retraces
the history of the VR theorem during the (semi-) analytic era of stellar
astrophysics. The first statement of what not much later became the
theorem appeared essentially \emph{en passant} in a research note of Heinrich Vogt
where he generalized Eddington's mass-luminosity relation. The `proof'
by Henry Norris Russell, which advanced the original claim to a theorem, was
enshrined later in Chandrasekhar's seminal reference work, \emph{An Introduction
to the Study of Stellar Structure}. After that the VR theorem lived
an apparently quiet and unquestioned life to the end of the 1950s when first
counterexamples, albeit rather un-astrophysical ones, were put forth.
The literature of the early 1960s contains more cases of by then more
naturalistic  star models that seemed to violate the VR theorem.
The following exposition covers the history up to the publication the
monograph of \citet{pss68} where the authors devoted a short 
chapter to the VR theorem and gave an impression of the thinking on this
matter at that time. 

A forthcoming second part of the history of the VR theorem will deal
with a revived interest in the topic in the 1970s, an epoch when
stellar evolution theory had turned into a computation-intensive 
enterprise with ever more complex input physics and ever more 
complicated resulting star structures.

\section*{First formulations}

The 1910s and 1920s were the years when astrophysicists started to
understand the stars as long-living self-gravitating fluid spheres.  Even
though the source of energy was not yet identified, the
thermo-mechanical structure of stars was already modeled mathematically. In
accordance with this intellectual achievement, the body of
history-related literature collecting, scrutinizing the contributions,
and cross-linking the players in the field is as huge as
authoritative. For those interested in the topic,
\citet{Hufbauer2006}, \citet{Gingerich1995}, \citet{Cowling1966}, and
references therein serve as fertile starting points.

For this marginalia it is sufficient to realize that in 1924 Arthur
Eddington published a paper with the title \emph{On the relation between
the masses and the luminosities of the stars} \citep{Eddington1924}
where he used his analytical thermo-mechanical star models to fit the
observed correlation of luminosities and masses of stars (see
Appendix~A for a discussion). Eddington collected the data of 46 stars
belonging to various kinds of binary stars for which masses and
absolute magnitudes were reported; all of them (plus five
pulsating variable stars) obeyed a remarkably smooth relation.
\begin{marginfigure}
\includegraphics{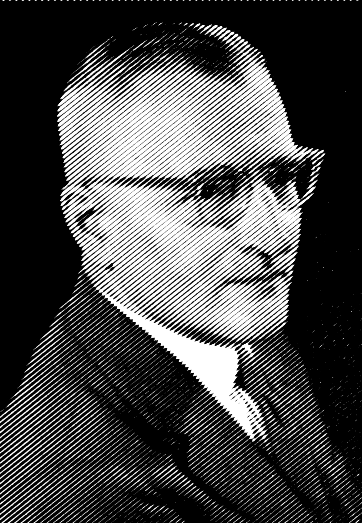}
\caption{Heinrich Vogt (1890 - 1968) after \citet{Bohrmann1968}}
\end{marginfigure}
In December 1925, Heinrich Vogt, an astronomer at Heidelberg's
K\"onigstuhl Observatory, submitted a short theoretical paper to the
\emph{Astronomische Nachrichten} \citep{Vogt1926} in which he
generalized Eddington's analytical star models by introducing spatially
variable forms of the mass-absorption coefficient
and of the energy generation rate. In
the last paragraph of the research note, Vogt mentioned laconically,
what he later referred to as the \emph{Eindeutigkeitssatz} (uniqueness
theorem):
\begin{quote}
  {[}...{]} \emph{Wir m\"ussen annehmen, da\ss\,die mittlere Dichte, die
  effektive Tempe\-ra\-tur und die absolute Leuchtkraft eines Sternes nur
  von seiner Gesamtmasse abh\"angen} {[}...{]}
\end{quote}
%-----------------------------------------------------------------------
\marginnote[-0.9cm]{Translation: We must assume that the mean density,
  the effective temperature and the absolute luminosity of a star
  depend on its total mass only.}
%-----------------------------------------------------------------------

Vogt allowed, additionally, for a small variation of the magnitudes of
the global stellar quantities at fixed mass because they might be
influenced by the nature of the stellar material, i.e. by the star's
chemical composition. In other words, Vogt claimed that the global
stellar quantities $\last(M,\mu), \teff(M,\mu),
\overline{\rho}(M,\mu)$
%-----------------------------------------------------------------------
\marginnote{The chemical composition, which is assumed to be
  homogeneous in the stars under consideration, be quantified by
  its mean molecular weight $\mu$.}
%-----------------------------------------------------------------------
are uniquely determined by the value of stellar mass and the star's
composition alone. As it
seems, the uniqueness claim was not particularly important to Vogt: About
one year after the first note on the subject, in a long review of the
theory of stellar structure and evolution, \citet{Vogt1927} did not
even touch the aspect of uniqueness of the solutions to the equations
describing the structure of the stars.  Another year later, when
\citet{Vogt1928} devoted a more extensive paper particularly to the
mass~--~luminosity law to emphasize once again that a form very close to
that of Eddington can be recovered even if the internal structure of
the stars differs from what Eddington assumed. And again,
the paper of 1928 does not mention the
\emph{Eindeutigkeits}-property.  Only in 1930, in the longest treatise
\citep{Vogt1930} concerning the relation of mass, luminosity, and
effective temperature of the stars~--~essentially an attempt to
understand the distribution of the stars in the Hertzsprung-Russell~(HR) 
diagram~--~did 
%-----------------------------------------------------------------------------
\marginnote[-2cm]{ At the time of Vogt's article, 
	the color/spectral-type~--~magnitude
  	diagram was typically referred to as the Russell diagram.
  	The designation `Hertzsprung-Russell' diagram
  	took over in the literature only in the second half of the 1930s.
  	\vskip 0.5cm}
%-----------------------------------------------------------------------------
Vogt refer briefly and informally to the dependence of the global
stellar quantities on mass and chemical composition; however, without
mentioning his 1926 paper. Even after the 1930 publication, one
wonders how much importance Vogt actually attributed to the uniqueness
conjecture and if he realized its consequences.

\medskip

%------------------------------------------------------------------ 
\begin{marginfigure} 
\includegraphics[scale=0.55]{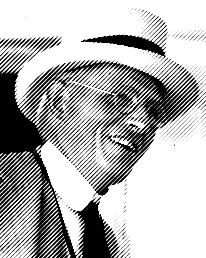}
  \caption{Henry Norris Russell (1877 - 1957) after
    \citet{devorkin2013}} 
\end{marginfigure}
%------------------------------------------------------------------ 
On the other side of the Atlantic, Henry Norris Russell studied the
physical basis of stellar evolution since the earliest days of modern
astrophysics \citep[cf.][]{DeVorkin2000}.  In the mid 1920s, he and
colleagues at Princeton Observatory overhauled the textbook
\emph{Astronomy} - \emph{A revision of Young's Manual of Astronomy},
updating it also with the latest ideas and results from the thriving 
field of stellar astronomy \citep{Russell1927}. In the second volume, in
Section~975, the authors state:

\begin{quote}
  {[}...{]} \emph{It is found that a star of given mass and composition
  will usually be in equilibrium for only one value of the radius, and
  hence for definite values of the luminosity and surface
  temperature. For stars of different masses these values will be
  different, but so long as the composition is the same, all the stars
  of a given luminosity will have to be of some one definite size,
  surface temperature, and spectral type.} {[}...{]}
\end{quote}
The above statement was made without further elucidation and without 
going into any technical details.  In any case, the uniqueness 
claim of Russell et al. 
appears more deliberately formulated and in particular physically more
coherent than Vogt's statement from a year earlier. In the same 
Section~975 of the textbook, the uniqueness property was
then applied to the interpretation of the structure of what later
became the HR diagram and of Eddington's mass~--~luminosity
diagram. Any scatter to the observed distribution of stars in these
diagrams that goes beyond the observational uncertainties was
interpreted by the authors to mean that {[}...{]} \emph{generation of heat}
{[}...{]} \emph{is different in different stars. The stars cannot therefore
all contain the same proportion of `active matter' }{[}...{]}, in
other words {[}...{]} \emph{they} [the stars] \emph{must differ in
composition}. Hence, in the textbook of \citet{Russell1927}, the
uniqueness statement of the structure of the stars is of auxiliary
use only, namely as a supporting argument in the interpretation of the
distribution of the stars in the HR diagram. 
The stars were thought to populate the HR plane as a function of
mass and of varying chemical composition. The mass was believed to
diminish with age so that stars evolve across the HR plane from high
to low mass during their life.  At the end of Section 975 of
\emph{Astronomy}, where contributors to the content of the
stellar-evolution discussion, including the uniqueness conjecture,
were referred to, only Russell himself and Eddington appeared
~--~Vogt's research note was not mentioned.

\section*{The path to theorem}

In a review, which discussed the state of the theory of the 
constitution of the stars, \citet{Russell1931} 
clarified that Vogt and, independently, he
himself formulated a few years earlier a theorem on the uniqueness 
of stellar structure. When referring to Vogt, Russell cited
the extensive paper of \citet{Vogt1930} on the nature of the
correlations of global stellar observables and their relation to their
internal structure rather than the short note of 1926.
One volume of the MNRAS later, \citet{Russell1931a} eventually set the records
straight: In an \emph{addendum,} he reported that Vogt brought to his
attention the proper reference containing the \emph{first }statement
of the uniqueness claim, which goes back to the year 1926.  In any
case, as early as 1931, Russell regarded the uniqueness statement as a
\emph{theorem }(of deep insight and as a contribution to the field of
stellar astrophysics which he considered to have remained much
undervalued in the community). Not much later, \citet{Russell1933}
reiterated his opinion in a non-technical survey on stellar
astrophysics in general, and stellar structure and evolution in
particular; he referred again to the \emph{Vogt theorem}
%---------------------------------------------------------------------
\sidenote{in part 2, p.~417}
%---------------------------------------------------------------------
as the [...] \emph{most important general proposition regarding stellar
constitution }[...].  As of then, Russell apparently attributed to
the uniqueness theorem much more importance than Vogt ever did in any
of his writing; this might be connected with Russell's interest to
thoroughly interpret the distribution of stars across the spectral
class~--~luminosity diagram, 
%--------------------------------------------------------------------------
%\sidenote{ i.e. in the HR diagram}
%--------------------------------------------------------------------------
on which he actively worked with the goal to pack all the stars into a 
coherent story in the framework of the theory of stellar evolution as it stood then
\citep[cf. ][]{Gingerich1995, Hufbauer2006}.

Russell, in contrast to Vogt, resorted already early on to a
mathematical argumentation and  \emph{proofed }the
uniqueness theorem \citep[cf.][]{Russell1931}: 
%----------------------------------------------------------------------------
\marginnote{In Section 10, Russell sets the stage by proclaiming 
self-confidently: \linebreak[4] [...]  \emph{The proof is simple.} [...]}
%----------------------------------------------------------------------------
The stellar structure problem was considered as the solution 
to a system of four differential equations with distributed boundary
conditions. Since the number of boundary conditions was counted to be three,
compared with four differential equations, Russell concluded after some
meandering that unique one-parameter sequences of solutions must exist for a 
prescribed chemical composition, and that without loss of generality, 
the star's mass can be chosen as this parameter.

Under the spell of the success of quantum theory, the young Danish
astrophysicist Bengt 
%-------------------------------------------------------------------------
Str\"omgren \sidenote{cf. \citet{Rebsdorf2003}} 
%-------------------------------------------------------------------------
set out, in the early 1930s, to improve the understanding of the HR diagram
adopting hydrogen-rich stellar models with the use of more elaborate
microphysics. In long review paper, \citet{Stroemgren1937} presented a 
comprehensive view of his understanding of the theory of the stellar 
interiors and of stellar evolution. A whole section
%---------------------------------------------------------------------------
\sidenote{Section II.16}
%---------------------------------------------------------------------------
of the exposition was dedicated to the uniqueness conjecture of 
Vogt and Russell.  Str\"omgren referred to it as the 
\emph{[...] Satz }
%---------------------------------------------------------------------------
\sidenote{\emph{Satz} in the mathematical sense of \emph{theorem}.}
%---------------------------------------------------------------------------
\emph{von Vogt und Russell [...]} (p.~477) and he also outlined its proof, 
following the line of argumentation of
\citet{Russell1931}. Not much later and apparently influenced by
Str\"omgrens work, the authoritative monograph  
\emph{ Stellar Structure} of \citet{Chandrasekhar1939} consolidated the
\emph{theorem} status of what, at best, should still have been considered
the \emph{conjecture} of Vogt and Russell. In Section VII.1 of his
book, Chandrasekhar contemplated, nonetheless, circumstances which
could void the VR theorem. 
He hypothesized material functions as sources of trouble; e.g. a nuclear
energy generation rate, which does not depend on the
\emph{local} values only of $\rho$ and $T$. Macroscopic
counterexamples, however, were yet beyond the intellectual horizon.

\citet{Kurth1953}
%----------------------------------------------------------------------------
\sidenote[][-1.0cm]{\emph{Kurth, Rudolf}, $^\ast$1917 in Germany,
  mathematician; doctorate (1948) and habilitation (1951) at the
  University of Berne, Switzerland, on topics of stellar dynamics.
  Later he pursued an academic career in England and the USA; he was a
  prolific writer of books on epistemology, philosophy, mathematics in
  physics and astronomy.}
%----------------------------------------------------------------------------
studied homology transformations of the stellar structure equations and
their properties (in a remarkably modern formulation). In this
context, Kurth formulated an aggravated version of the VR theorem in
that he progressed from \emph{if} to \emph{iff}: In the framework of
homologous, chemically homogeneous star models in complete equilibrium
%---------------------------------------------------------------------
\sidenote{Complete equilibrium in the sense of \emph{thermal } and
  \emph{hydrostatic } equilibrium.}
%---------------------------------------------------------------------
he concluded that the stars have nearly the same internal structure
iff they have nearly the same mass and nearly the same chemical
composition.  Only a few sentences after this conclusion, Kurth
cautioned the reader that 
%-----------------------------------------------------------------------------
\marginnote{Translation: Nothing is proved, these are all plausibility
considerations only }
%-----------------------------------------------------------------------------
[...]  \emph{Bewiesen ist nichts, es handelt sich nur
  um Plausibilit\"atsbetrachtungen} [...],
all statements were indeed clearly declared as plausibility
considerations; in particular, he explicitly assumed that solutions to
the stellar structure equations exist. Finally, Kurth also
distinguished between the pure stellar structure problem
(i.e. stationary solutions to the structure equations), for which the
VR theorem was formulated, and the full stellar-evolution problem
which, by its very nature, is a time-dependent problem.  Therefore, he
warned that any extrapolation of conclusions from the
application of the original VR theorem to the realistic
stellar-evolution problem has to be treated with utmost caution.

Only a few years later, \citet{Odgers1957}
%----------------------------------------------------------------------------
\sidenote[][-0.2cm]{\emph{Odgers, Graham J.}, $^\ast$1921 in
  Australia; he spent his life as an astronomer in Canada where he started
  his career with theoretical and observational research. 
  Later, he focused on instrumentation, in particular the construction 
  of large telescopes. On the Canadian side, Odgers was 
  instrumental in the construction of the CFHT on Mauna Kea.}
%----------------------------------------------------------------------------
reported his attempts to construct homologous series of chemically
homogeneous star models in complete equilibrium. The aim was to
derive a simple analytical formulation of the mass~--~luminosity law
of main-sequence stars.  Already early on in the paper, Odgers
criticized the mathematical assumptions that entered the uniqueness
statement of the VR theorem and he even deduced an explicit homologous
series which violated it. Because the inferred energy-generation law
in the conflicting  model series was unphysical in the stellar 
context the counterexample to the VR theorem was considered 
at best of formal interest. Even though Odgers' paper was an 
actual mathematical blow for the VR \emph{theorem,} 
the paper made no impact in the astrophysical community:
%-------------------------------------------------------------------------------
\marginnote[-0.6cm]{At the time of this writing, an ADS query returned 6
  citations to the report; all between 1972 and 1978.}
%-------------------------------------------------------------------------------
The circumstances under which multiple solutions occurred were either
stellar-physically unacceptable or they were so restrictive that their
realization in nature seemed unlikely. A further handicap of the
Odgers paper was that it was published and circulated as an observatory
report only and as such must have had a diminished audience. Last but not
least, the report's content is very formal and likely was too tough to digest 
for many in the astronomical community.

In the second edition of his textbook, \emph{Aufbau und Enwicklung der Sterne},
\citet{Vogt1957} came back to the
\emph{Eindeutigkeitssatz der Theorie des Sternaufbaus} and devoted a whole
chapter to is. As in earlier publications, Vogt did not adopt 
a particularly mathematical point of
view to discuss the problem. Instead, he merely insisted that the
structure equations admit of unique solutions as long as the material
properties (such equation of state, opacity, or nuclear energy generation)
are well defined functions of thermodynamic state variables.
As means to destroy the
one-parameter families (characterized by different chemical
composition) of evolutionary tracks on the HR plane, Vogt contemplated 
physical effects such as rotation, electromagnetic braking, 
or tidal effects in binaries.

\medskip

Despite the lack of evidence of any impact in stellar astrophysics of
the report of \citet{Odgers1957} his paper marks the begin
of the era in which multiple solutions began to pop up in the 
ever more detailed stellar-model computations. For example,
\citet{Cox1964} calculated low-mass pure helium star models in 
complete equilibrium.  Below a critical mass, $M_{\mathrm{min}}$, 
helium stars cannot maintain steady helium burning. Around this
$M_{\mathrm{min}} \approx 0.305 \msol$,
a low- \emph{and} a high-density solution with the same total
mass and with identical composition were revealed. Not much later, for pure carbon
stars too, double solutions for equal-mass model stars were encountered
\citep{Deinzer1965}. For carbon stars,
%---------------------------------------------------------------------
\marginnote{No neutrino losses: $M_{\mathrm{min}}\approx 0.7
  \msol$, including neutrinos: $M_{\mathrm{min}}\approx 0.82 
  \msol$ for steady carbon burning.
}
%---------------------------------------------------------------------
the minimum mass is larger than that of the helium stars.
Neither in \citet{Cox1964} nor in \citet{Deinzer1965} is there any
indication that the authors connected the double solutions with a
failure of the VR theorem. \citet{Bodenheimer1966} on the other hand
questioned the validity of the VR theorem upon realizing that 
his pre~--~main-sequence model stars all converged
essentially to the same evolutionary locus along the Hayashi line,
independent of the initial conditions he prescribed for his model
sequences. Even though
Bodenheimer was the most attentive author back then, his models do not
serve as counterexamples to the VR theorem because they are
\emph{not} in complete equilibrium as they need to be for the orginal VR theorem to be applicable.

A next higher level of complexity in stellar modeling was reached with
composite models that consisted of a core and of a grafted envelope, both
in complete equilibrium but both with a differing chemical
composition. Adopting the mass of the core as the
control parameter allows to study the dependence of the physical
properties of a \emph{series }of star models under continuous variation of
the control parameter, such model sequences are known as \emph{linear
series}.  To investigate the onset of the Sch\"onberg-Chandrasekhar
instability, \citet{Gabriel1967} chose the relative mass,
$q_{\mathrm{He}}$, of the inert helium core below a hydrogen-burning 
shell and a hydrogen-rich envelope as the control parameter of their linear series of model stars with constant total mass . 
%---------------------------------------------------------------------------
\marginnote[-1.3cm]{N.B. Increasing the magnitude of $q_{\mathrm{He}}$
  can be understood as an emulation of stellar evolution through a
  sequence of equilibrium states.}
%---------------------------------------------------------------------------
Gabriel and Ledoux concluded
that the instability develops at a turning point of their linear
series. In the neighborhood of this turning point, the stellar 
structure equations were observed to admit of double solutions at
constant $q_{\mathrm{He}}$. Investigations of the stability of these
double solutions revealed then that one branch was secularly unstable. 
Even though double-solutions for the same
stellar mass and the same chemical-composition profile were encountered,
the result was not yet discussed in the context of the VR theorem. The
situation changed when \citet{Gabriel1968} studied pure carbon stars
in the neighborhood of the respective $M_{\mathrm{min}}$, again they 
found that only one branch, the
low-degeneracy one, of the double-solution region was secularly
stable. Eventually, the authors concluded that turning points 
of linear series signal violations of the classical VR theorem. 
Resorting to a more restricted formulation, \citet{Gabriel1968} tried to save 
the VR theorem by adding the aspect of secular stability:
[...] \emph{For a given mass and chemical composition there exists only one
  secularly stable configuration. }[...]

\medskip

Consulting
%-------------------------------------------------------------------------
\citet{pss68}\sidenote[][-0.5cm]{The monograph, \emph{Principles of Stellar
    Structure}, is referred to as PSS subsequently.},
%------------------------------------------------------------------------
who published a comprehensive textbook which details knowledge and understanding 
of structure and evolution of 
%-------------------------------------------------------------------------
simple \sidenote{Simple in the sense of radial symmetry of the star
  models, devoid of rotation and magnetic fields.} 
%-------------------------------------------------------------------------
single stars by the mid 1960s, one finds it to offer indeed an appropriate
endpoint to the first part of this review of the history of the VR theorem. 
In PSS, a whole~--~albeit 
%------------------------------------------------------------------------
short~--~chapter % \sidenote{Chapter 18} 
%------------------------------------------------------------------------
is devoted to the VR theorem; that choice met criticism already
early on by one of the reviewers of the books \citep{Sweet1969} and it 
likely sheds light on the importance attributed to the VR theorem at that
time.  Be it as it may, Chapter 18 of PSS is very useful here because
it offers a glimpse at the perception of the VR theorem in the mid 1960s. 
Early on in the discussion, the authors emphasized that the VR theorem 
is not a theorem in a strict sense because cases of multiple solutions
had been encountered
%-----------------------------------------------------------------------
\marginnote{The known double-solutions were either rather abstract and
  of little relevance to the Universe or one of the
  double solutions was secularly unstable and would not survive for
  long in nature. }
%-----------------------------------------------------------------------
and that a watertight proof had never been put forth. Nonetheless,
Cox and Giuli could not resist the temptation to give a kind of
a plausibility-`proof' of the VR theorem, following the
line of thought already present in \citet{Russell1931}. The
system of equations that entered the proof remained those of a stellar
configuration in complete equilibrium so that the problem reduced to
system of ordinary differential equation. The separated boundary
conditions were introduced and it was argued, without going into any mathematical
detail, that the implied algebraic constraints of the boundary
conditions, being of lower dimensionality than the dimension of the
solution space, ensure [...]\emph{ under ordinary conditions}~[...]
unique solutions. More originally, Cox and Giuli offered also a physical
interpretation of the VR theorem, they resorted to an 
order-of-magnitude discussion of the stellar structure problem 
(PSS, chapter 18.2) and thereby collapsed the differential equations
to a set of algebraic relations.  They showed that if
pressure, density, temperature, and radiated luminosity were prescribed,
all emerging relations can be expressed as functions of mass, radius, and
chemical composition. Upon prescribing  additionally also thermal
equilibrium, the radius dependence of the set of algebraic equations
can be eliminated so that eventually the order-of-magnitude
approximations of the physical quantities of a star in hydrostatic and
thermal equilibrium are found to depend on mass and chemical
composition only. Although intuitively attractive, the method applied only
to a model star in a coarse integral sense and failed to be
mathematically rigorous (cf. Appendix B).

\medskip 

Astrophysics is not mathematics; in the latter, the VR theorem would
have met its fate once \emph{one }counterexample popped up~--~independent of how
academic it were. The astrophysical community, on the other hand, put up with
the dilemma of the VR theorem and its counterexamples. After all,
in astronomically relevant cases it seemed to remain predictive and
explanations sounded plausible. Nonetheless, the beginning era of
ever faster and easier accessible electronic computers at the end of the
1960s allowed to compute physically complex models in large
numbers and the rapidly growing repository of stellar models had 
intriguing challenges in store.

The forthcoming second part of this marginalia on the history of
the VR theorem will focus on the developments in the 1970s when 
a few astrophysicists set out to look more
closely into matter of the VR theorem after ever more complicated star
models could be computed and some of them  exposed violations of the VR
theorem. The field benefited from fresh insights imported by people
who applied to stellar astrophysics heavier mathematical machinery than usual.

\bigskip

\section*{Appendix A: Eddington's mass~--~luminosity
  relation} \label{sec:Appendix:-Eddington's-mass}

A correlation of stars' masses and luminosities was hinted at as
early as 1911; it was first mentioned in a clause of a paper on the motion
of the stars in the Galaxy \citep{Halm1911}. 
Analyzing an appropriately chosen sample of 14 binary stars, \citet{Hertzsprung1923} 
was able to report quantitatively on a relation of mass and brightness of his sample 
stars, finding clear evidence that more
massive ones were consistently brighter than the less massive brethren.
It was Eddington, however, who pushed the issue of the
mass~--~luminosity relation ($M-L$ relation in the following) further
because he relied on it as an observational foundation on which he could rest his 
mathematical modeling of the internal structure of the stars
\citep{Eddington1924}. 

\begin{figure*}
  \includegraphics[width=0.46\linewidth]{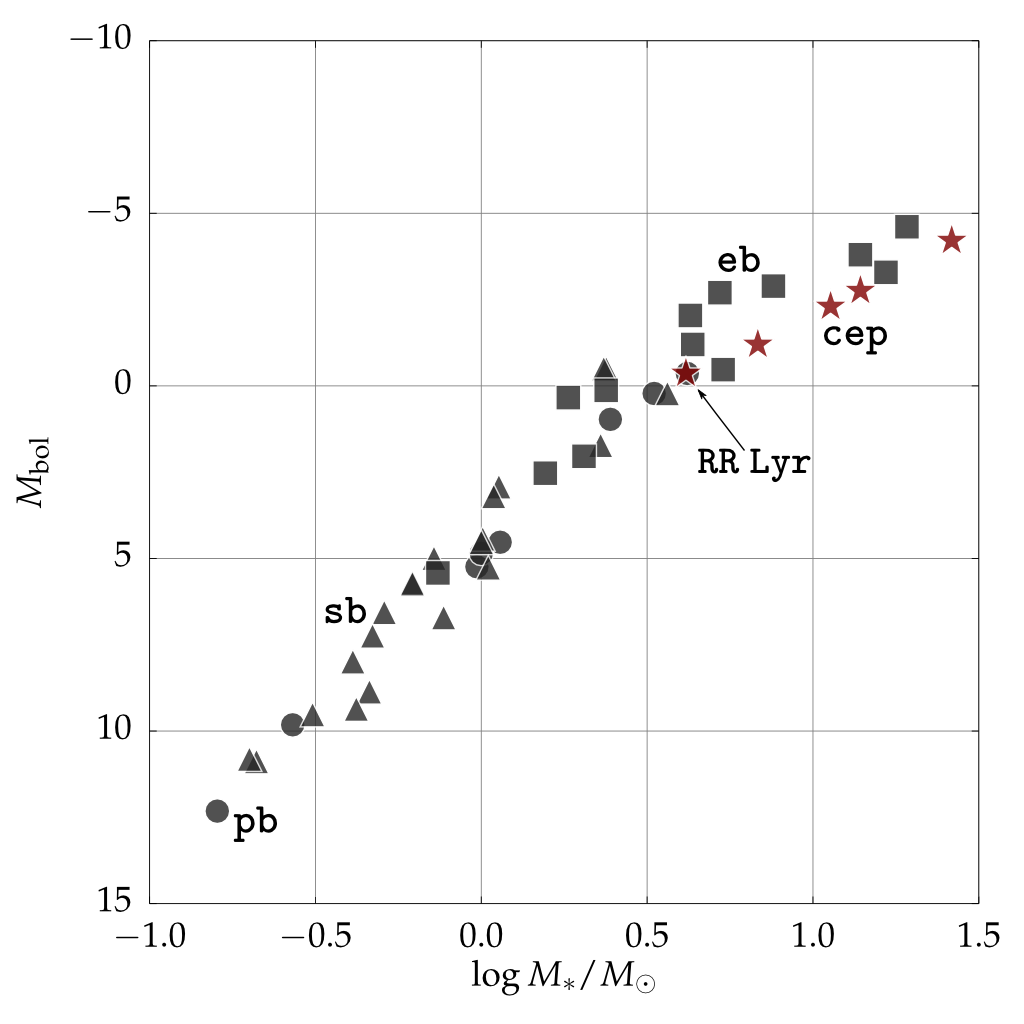}
  \hspace{0.3cm}
  \includegraphics[width=0.46\linewidth]{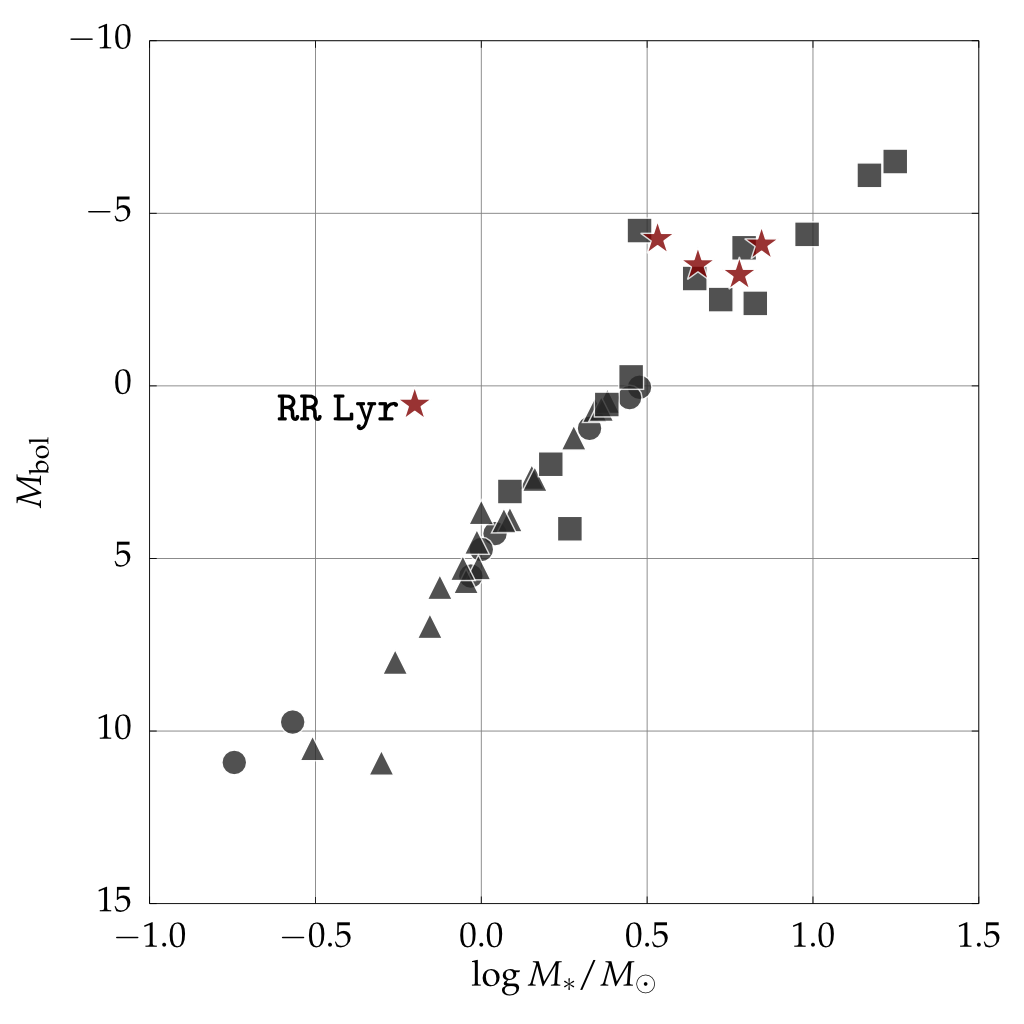}
  \caption{Mass-luminosity relation as used in Eddington's studies
    in the early 1920s. The left panel contains the stars with the
    calibration known to \citet{Eddington1924}. The right panel
    shows the relation when using current calibrations of the same 
    objects. (Data harvested via SIMBAD and ADS).
    \label{fig:ML_Eddington}}
\end{figure*}

Figure \ref{fig:ML_Eddington} displays the same data as were shown in
Fig.1 of \citet{Eddington1924}. The different markers in the plot
identify different classes of calibrated stars; filled circles show
the \texttt{pb} stars, the primary binary stars, triangles are the
\texttt{sb}, the secondary binaries, squares show the \texttt{eb}, the
eclipsing binaries, and finally, the asterisks stand for \texttt{cep},
the pulsating variables. Eddington referred to them collectively as
Cephe\"ids; this despite the fact that the star RR Lyr  was also in the
sample. The physical differences between Cephe\"ids and RR Lyrae 
variables were not known at the time.

From the present point of view, Eddington's adding pulsating stars, the \texttt{cep} group,
to the graph to make the case of an $M-L$ relation is a 
dubious undertaking. It was essentially pure luck that made the outcome 
to look so seemingly convincing. At the time of Eddington's article, no direct
determinations of masses of Cephe\"ids were available. For the stars to
find their place in the figure, Eddington resorted to his theoretical
$M-L$ relation and applied it ad hoc to the Cephe\"ids too. The computed
stellar mass was then iterated until the pulsation period of the 
\emph{modeled }Cephe\"id eventually converged to the observed period. 
Therefore, the pulsating stars in Fig.~\ref{fig:ML_Eddington} 
are no fundamental indicators of the existence an $M-L$ law but the 
result of an already plugged-in $M-L$ relation coupled with the pulsation theory
of \citet{Eddington1919}. In contrast, the derivation of the masses 
of the members of the binary-star sample relies
on Newton's laws acting in a $1/r$-gravitational potential only.
Therefore, only the binary stars serve as legitimate indicators 
of the correlation between mass and luminosity. 

An inspection of the right panel of Fig.~\ref{fig:ML_Eddington} 
makes clear that correlation of $M-L$ relation of the binary 
stars persists also with modern physical calibrations of the binary stars, 
although the scattering increases at very low and very high luminosities.  
Furthermore, the relation based on modern data has a steeper slope plus a 
few outliers in the graph; the Cephe\"ids in particular scatter.
Concentrating on the arguably small number of Cephe\"ids with 
the equally small mass spread, the modern data
do not really call for the same $M-L$ relation as it is suggested by
the mostly main-sequence binary stars. Nonetheless, the overall
relation is still impressively tight, particularly when accounting for
the fact that very different kinds (evolutionary stages) of stars convene in the graph.

Because the stars spend most of their lifetime buring hydrogen most
stars observed in the sky are therefore likely in their main-sequence
phase. Along the main sequence, the stars' luminosities grow with
increasing mass. One representative form of an empirically calibrated
main-sequence $M-L$ relation is e.g. from \citet{Smith1983}:
\[
  \log \llsol = 4.0\cdot\log \mmsol 
  \,\,\,\,\,\,\mathrm{for}\,\,\mbox{\ensuremath{\mmsol>0.43}}.
\]
The binary-star data entering Eddington's $M-L$ relation fit the above
relation quite well so that we can assume that the respective stars 
are indeed likely in their main-sequence phase of their life.

The Cephe\"ids, on the other hand, are as we know today radially pulsating
intermediate-mass supergiants. Under favorable circumstances
intermediate-mass stars loop across the HR diagram during their
central helium-burning stage and some of these looping stars migrate
through the classical instability strip to become then observable as
Cephe\"ids. From stellar-evolution modeling we learned that close to the
instability strip the luminosities of the same branches of these blue loops  
tend to be ordered in mass; i.e. Cephe\"ids of different
mass but a comparable evolutionary stage tend to obey  
an $M-L$ relation too. The $2^{\mathrm{nd}}$ crossing of the instability strip during 
the early core helium-burning phase is usually the slowest 
and therefore the favored one to observe Cephe\"ids in.
Adopting hence this second crossing as the relevant one here, 
a fit to the computed intersections of evolutionary tracks with 
the instability strip \citep{Chiosi1993} reads as
\[
   \log \llsol = 3.57\cdot\log \mmsol + 0.54\,.
\]
Interestingly enough, the slopes of the Cephe\"ids' and the
main-sequence stars' $M-L$ relation happen to be quite similar. From
all we know, this is \emph{an accident of nature}.  The vertical
displacement, $\Delta\log(\llsol)$, of the two relations say at $\mast
= 5\, \msol$ is only $0.23$ so that the composite nature of the
observed $M-L$ relation in Fig.~\ref{fig:ML_Eddington} is hardly
discernible, in particular in the presence of unavoidable scattering
of observational data.

Apart from some increased scattering in the modern version of
Eddington's $M-L$ relation, introduced by the pulsating variables, one
obvious disagreement is apparent in comparison with the original one:
The data point of the star RR~Lyr lies far off the general trend. 
This is no surprise: Today we know that the family of RR Lyr variables 
is made up of low-mass ($\approx 0.6 \msol$) population~II stars living on the
horizontal branch. Rather than following an
$M-L$ relation of the above kind, the spread in mass and in luminosity
is small so that the \emph{class} of RR Lyrae variables  
would form kind of a clump around the isolated prototype RR Lyr in 
Fig.~\ref{fig:ML_Eddington} (right panel).

\medskip
\section{Appendix B: The nature of the equations \label{sec:Appendix-B: Uniqueness}}

To properly state what astrophysicists mean if they talk stars on the theory level,
the set of the governing equations and assumptions are laid out in the following.
In each case of the following collection of formulae, the first line contains, as
the starting point, the general fluid-dynamical equations in their
Lagrangian form. We follow mostly the notation of \citet{mimi}
(only if not self-evident, deviations therefrom are explained). The second line
specializes then on spherical
%-------------------------------------------------------------------------
symmetry\sidenote[][-3cm]{Spherical symmetry is appropriate for non-rotating,
  non-magnetic star. Even though it seems intuitively obvious that a
  static self-gravitating fluid configuration assumes the form of a
  sphere, the proof that the sphere is the \emph{only}
  equilibrium figure came relatively late, see e.g. \citet{Poincare1902} 
  with a proof which relied on Lyapunov's master thesis of 1884 or consult
  \citet{Carleman1919} who used a more geometrical ansatz.}
%-------------------------------------------------------------------------
and on mass as the independent variable~--~choices usually adopted in
stellar structure/evolution modeling : \marginnote[+0.5cm]{Continuity
  equation} \marginnote[+0.3cm]{Mass equation}
\begin{align}
\label{eq:mass}
\Diff_t \,\rho &= - \rho\,\left(\vec{\nabla}\cdot \vec{v}\, \right) \, ,\\
\Diff_m\,r     &= \frac{1}{4\pi r^2 \rho} \, .
\end{align}

\marginnote[+1.4cm]{Poisson equation (elliptic equation)}
\begin{align}
\label{eq:poisson1}
\Delta \Phi &= 4 \pi \rho G     \, ,   \\
\label{eq:poisson2}
          g &= -\frac{G m}{r^2} \, ,
\end{align}
In spherical symmetry, eq.~\ref{eq:poisson1} can be integrated once; defining 
$g \doteq - \diff_r \Phi$ leads then to eq.~\ref{eq:poisson2}.

\marginnote[+1.3cm]{Cauchy's equation (hyperbolic equation)}
\begin{align}
\label{eq:cauchy}
 \rho \Diff_t\vec{v} &= \vec{f} + \vec{\nabla}\cdot\tens{T} \, , \\
 \Diff_m P           &= -\frac{Gm}{4\pi\,r^4} - \frac{1}{4\pi\,r^2}\,\Diff_t^2r \, .
\end{align}
The stress tensor $\tens{T}$ is made up of the components of viscous stress
and it harbors on its diagonal the isotropic hydrostatic pressure. Neglecting viscosity turns Cauchy's equation into Euler's equation. 
Physically, the stress tensor is relatively unimportant in
stellar matter; for numerical reasons (in `hydrodynamical' computations) an
artificial stress tensor might become instrumental to smear out shocks
and to stabilize the computational scheme.

The change of the specific heat content of the stellar matter, $q$, 
remodeled with the help of the $1^\mathrm{st}$ law of thermodynamics, 
can be written as: 
\marginnote[+1.7cm]{Gas energy equation}
\begin{align}
\label{eq:energy}
\rho\,\Diff_t q &= \rho \left[D_te + P\,\Diff_t\left(\frac{1}{\rho}
                                               \right)\right] =
                                               \Psi_{\mathrm{visc}}-\vec{\nabla}\cdot
                                               \vec{F} + \rho\,
                                               s_{\mathrm{nuc}} \, ,   \\
\Diff_t q &= - \Diff_m L + \varepsilon  \, .
\end{align}
%The specific internal heat (i.e. internal heat per mass unit), $q$,
%can be related to the specific entropy, $s$, by means of the first law
%of thermodynamics, i.e. $T \Diff_t s = \Diff_t q$, so that the gas
%energy equation measures the change of stellar matter's entropy. 
As in the momentum equation, the dissipation term~--~here the viscous
energy dissipation function $\Psi_{\mathrm{visc}}$~--~is physically
not important in stellar interiors except if sharp fronts develop and
then in connection with the necessity to stabilize the numerical
treatment. In canonical quasi-static stellar-evolution computations,
$\Psi_{\mathrm{visc}}$ can be neglected.

The energy flux is denoted by $\vec{F}$. Astronomers prefer to work
with the local radially streaming luminosity, which is defined as $L=4\pi r^2 F_r$, with $F_r$ being the radial component of the local energy flux. Furthermore, the
energy source, $s$, is attributed to nuclear burning in the star, therefore it
is subscripted with `nuc'. In the stellar astrophysical
form of the equation, as shown on the second line, the nuclear energy
input rate, measured as energy per unit time and unit mass, is
referred to as $\varepsilon$. It is important to keep in mind that
\emph{energy gain} by nuclear burning contributes positively and
possible \emph{energy loss} by neutrino production is to be subtracted because
neutrinos do not contribute to the heat content of the stellar
%--------------------------------------------------------------------------------
matter.\marginnote[-0.9cm]{Extreme conditions such as encountered e.g. 
during a stellar core collapse can trap even neutrinos and hence require then 
a careful treatment of the energy budget \emph{including }the neutrinos. }
%--------------------------------------------------------------------------------

To get a handle on the physics of the energy flux, stellar astrophysicists adopt
Fourier's law to model the flow of photons as a diffusion process driven by the
spatial temperature gradient. Even energy transport by material motion can
be appropriately accommodated. 
\marginnote[+0.8cm]{Fourier's law for the flux (parabolic equation)}
\begin{align}
\label{eq:diffusion}
\vec{F}   &= -K\cdot\vec{\nabla} T  \, ,               \\
\Diff_m T &= -\frac{G m}{4\pi r^4 P}\cdot\nabla_0 \, . 
\end{align}
The quantity $K$ denotes the coefficient of thermal conductivity; in
radiative regions it can be written as $K= ac T^3 /(3\kappa \rho)$
with $\kappa(\rho,T,\vec{\chi})$ being the Rosseland opacity.
The quantity $\nabla_0$ measures the temperature stratification: 
\begin{equation*}
\nabla_0 = \frac{\diff \ln T}{\diff \ln P} = 
                             \begin{cases}
                               \,\nabrad          & \text{radiative region,} \\
                               \,\nabla_{\text{c}}  & \text{convective region.}
                             \end{cases}
\end{equation*}
with $\nabrad = 3 \kappa L P / (16 \pi a c G m T^4)$ being a purely
local function of stellar quantities. In the easiest case of a 
stellar-convection description, 
such as in elementary mixing-length models for example, also $\nabla_{\text{c}}$ is a
function of local variables alone. More elaborate treatments of
convection can, however, introduce non-local contributions.

Nuclear burning is the source of a star's evolution; the resulting
spatio-temporal change of nuclear species $X_i$ 
%-------------------------------------------------------------------------
\marginnote{$X_i$ be the mass fraction of nuclear species $i$; 
			$\sum_i X_i = 1$.}	%-------------------------------------------------------------------------
in a star can formally be written as 
%-------------------------------------------------------------------------
\marginnote[+0.3cm]{$i \in \left[1,\dots,N_{\mathrm{spec}}
  \right]$, with $N_{\mathrm{spec}}$ the number of species accounted
  for in the stellar matter.}
%-------------------------------------------------------------------------
\begin{equation}
\Diff_t X_i = Q_i - S_i + \vec{\nabla}\left(\sigma_{\mathrm{D}}\vec{\nabla}X_i\right)\,.
\end{equation}
Apart from the source-, $ Q_i$, and the sink-term, $S_i$, both
determined by the type and complexity of the nuclear burning network,
nuclear species can be smeared out spatially by a multitude of
physical processes (such as convection, thermohaline mixing,
semi-convection, settling, levitation); these transport processes are
hidden away in a diffusion-type term in the equation with the
particular physical process manifesting itself in the specification of
the diffusion coefficient $\sigma_{\mathrm{D}}$. In the majority of
the numerical realizations, the computation of the nuclear evolution
is decoupled from the stellar structure 
%-----------------------------------------------------------------------------
problem.\sidenote{For a modern approach \emph{coupling} nuclear
  with stellar-structure evolution, see Appendix B of  \citet{Paxton2013}}
%-----------------------------------------------------------------------------
We presume that at each epoch $t$, the vector $\vec{\chi}\doteq
\left(X_1(m,t),\dots , X_{N_{\mathrm{spec}}}(m,t)\right)$ is known via
some suitable computational procedure.

\emph{Boundary conditions} for the stellar structure equations are
distributed ones with the natural choices in the \emph{center}: $r=0$ and
$L=0$ at $m=0$. The \emph{surface} is, by its very stellar nature,
ill-defined and requires suitably chosen physical approximations: 
Traditionally popular is the assumption of thermal equilibrium of
radiation and matter fields at the photosphere leading to: 
$L=4\pi r^2 \sigma T^4$ at $m=M_\ast$. The radius at the
photosphere is then set equal to star's radius $r=R_\ast$, and the
temperature at the photosphere corresponds to the so called effective
temperature $\teff$. The second outer boundary condition, a mechanical one, 
determines e.g. the pressure at the photosphere:
$$
  P = f(\rho,T,\kappa(\rho,T,\vec{\chi}))\, ,
$$
with some suitable function $f$, which approximates the type of
atmosphere which exerts its pressure, $P$, on the photosphere. For simplicity's
sake, and likely sufficient for pure mathematical considerations, it
is sufficient to assume some ad hoc constant pressure at the photosphere:
$$
 P = P_{\mathrm{phot}}=\mathrm{const.} \ll P_{\mathrm{center}} \,.
$$

Finally, \emph{initial data} that specify the state of the star's 
structure are required to get a model sequence started in time. 
Typically, such a time evolution
is initialized with some simplified star model in hydrostatic and 
if possible also in thermal equilibrium. Both assumptions ensure
that pressure and temperature are continuously differentiable in
space. Density, on the other hand, can have discontinuities, depending
on the spatial structure of the composition vector;
luminosity will also react accordingly. Think, for example, of an
initial model with a pure helium core and a pure hydrogen
envelope: Across the H/He interface pressure and temperature are
continuous whereas density and, at sufficiently high temperature, also
luminosity develop finite jumps.

\medskip

Where astronomers were apparently too light-hearted in `proofing' the VR theorem,
mathematicians, on the other hand, in particular
those rising a warning finger, such as \citet{Kurth1953}, were essentially absent.
The asterophobia of the mathematicians is likely caused by the fact
that the equations that model the structure of the stars cannot be
pigeon-holed: Depending on the specific simplifications introduced 
to the system of structure and evolution equations, they can change 
the mathematical character so that different mathematical tools must 
be applied to the formal study the problem; on the other hand 
also different numerical methods must be implemented to 
tackle the computational problem. The simplest approach to model stars, 
which likely has canalized early thinking of proofing the VR theorem, 
namely the one used to compute polytropes is shortly touched upon in the following.

\newthought{Separating mechanical and thermal parts} of the stellar structure 
problem was the first ansatz to come to grips with understanding the interior 
conditions of stars. To establish the necessary barotropic conditions, frequently a
%--------------------------------------------------------------------------
\emph{polytropic}\sidenote[][-1.2cm]{The relation $P\propto \rho^{1+1/n}$, with $n$ being the polytropic index, constitutes a stratification relation rather than a state relation within a prescribed fluid element.} 
%--------------------------------------------------------------------------
relation was postulated. In the static case, i.e. in absence of a velocity field, 
eqs.~2, 4, and 6 morph into the 
venerable Lane-Ritter-Emden equation. In the formative years of theoretical
astrophysics, this equation was solved as an initial-value 
problem (IVP): The computation started in the regularly singular center of 
the model with prescribed values of the dependent variable and its
derivative. The integration was followed out to the first root of the dependent
variable; its location was then adopted as a measure of the radius of the model
star. Looking at the problem as an IVP ensured existence and
uniqueness of the solution by the sufficiently smooth character of the 
right-hand side of the ODE via Picard-Lindel\"of's theorem. 

More generally, in particular with a
non-vanishing velocity term, the original mechanical fluid-dynamical
equations (eqs. 1, 3, and~5, closed with a polytropic relation between
$\rho$ and $P$) constitute the so-called \emph{Euler-Poisson} problem.  
Assuming a compact support for the density and hence for the whole problem,  
i.e. $\rho >0$ obtains for a finite volume only and defining the outer
boundary  of the gravitationally-bound fluid sphere by $\rho(R) = 0$ bring
about considerable mathematical complications; a substantial body of
literature exists on uniqueness and evolution of such boundary-value
problems (BVPs). For some recent advances, consult e.g. \citet{Deng2003} 
who proofed uniqueness theorems for the static case, i.e. the
BVP which must be solved for the structure of polytropic spheres. 

Even though the proof of the VR theorem in \citet{Russell1931} (and all later
repetitions thereof) refers to the stellar-structure problem as a BVP, the 
presentation of how the equations are solved is reminiscent of the
direct integration of the Lane-Ritter-Emden equation. In other words, from 
reading \citet{Russell1931} one comes away with the impression that existence and 
uniqueness  properties were implicitly influenced by the IVP experience 
gained with polytropes. However, existence and uniqueness statements
for higher-order BVPs are mathematically formidable; we
superficially referred to the easiest case, 
that of the Euler-Poisson system, just before.   
    
\bigskip

Properties of the equations of more realistic approximations to the
stars' structure and their evolution will be a topic in the second part of 
this essay on the history of the Vogt-Russell theorem.  

\newpage

\newthought{Acknowledgments}: NASA's Astrophysics DataSystem, for the
old literature, and CDS/Simbad, pointing to the literature with modern
astrophysical calibrations of Eddington's $M-L$ relation star-sample,
were used extensively. Thanks to H.H. for hospitality at the dimly 
lit kitchen table where much of this marginalia crystallized.

\bibliographystyle{aa}
\bibliography{StarBase}

\end{document}